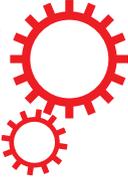

# SCIENTIFIC REPORTS

**OPEN**

# Sleep-like slow oscillations improve visual classification through synaptic homeostasis and memory association in a thalamo-cortical model

Cristiano Capone[1], Elena Pastorelli[1,2], Bruno Golosio[3,4] & Pier Stanislao Paolucci[1]

The occurrence of sleep passed through the evolutionary sieve and is widespread in animal species. Sleep is known to be beneficial to cognitive and mnemonic tasks, while chronic sleep deprivation is detrimental. Despite the importance of the phenomenon, a complete understanding of its functions and underlying mechanisms is still lacking. In this paper, we show interesting effects of deep-sleep-like slow oscillation activity on a simplified thalamo-cortical model which is trained to encode, retrieve and classify images of handwritten digits. During slow oscillations, spike-timing-dependent-plasticity (STDP) produces a differential homeostatic process. It is characterized by both a specific unsupervised enhancement of connections among groups of neurons associated to instances of the same class (digit) and a simultaneous down-regulation of stronger synapses created by the training. This hierarchical organization of post-sleep internal representations favours higher performances in retrieval and classification tasks. The mechanism is based on the interaction between top-down cortico-thalamic predictions and bottom-up thalamo-cortical projections during deep-sleep-like slow oscillations. Indeed, when learned patterns are replayed during sleep, cortico-thalamo-cortical connections favour the activation of other neurons coding for similar thalamic inputs, promoting their association. Such mechanism hints at possible applications to artificial learning systems.

Human brains spend about one-third of their life-time sleeping. Sleep is present in every animal species that has been studied[1]. This happens notwithstanding two negative facts: the danger caused by sleep, that diminishes the capability to defend from predators and other threats, and the reduction of time available for activities targeting immediate rewards (e.g. hunting or gathering food). Having survived the evolutionary selection in all species, sleep must therefore provide strong advantages. Another notable fact is that newborns' human brains occupy the majority of their time asleep, nevertheless they learn at a very fast rate. For this and other motivations for studying sleep, also in relation to consciousness, see e.g.[2]. Moreover, even if occasional awakening does not seriously impair brain biology and cognitive functions, in the long term chronic deprivation of sleep produces measurable effects on cognition, mood and health[3]. Experimental studies like[1,4] investigated the effects of sleep on firing rates and synaptic efficacies. In[1] the authors formulated hypotheses about homeostatic processes occurring during sleep. A possible motivation for the brain entering in the sleep activity would be to set a better energy consumption regime during next wakefulness cycle. This might be obtained by reducing firing rates and the amplitude of evoked post-synaptic potentials. It could be reached by pruning non necessary synapses and by reducing synaptic weights within the limits imposed by the conservation of adequate coding for memories. In[4] Watson et al. propose a novel intriguing experimental evidence. They used large-scale recordings to examine the activity of neurons in the frontal cortex of rats and observed that neurons with different pre-sleep firing rate are differentially modulated by different sleep substates (REM, non-REM and micro arousal). Sleep activity such as slow waves activity and sharp-waves ripples have been shown to be beneficial for memory consolidation[5,6] and task performances optimization[7]. A few computational models have been developed to investigate the interaction of sleep-like activity and plasticity. In[8], the authors showed that Up states specifically mediate synaptic down-scaling with beneficial effect on signal to noise ratio received by post-synaptic neurons. The effect of thalamo-cortical sleep on pre-stored time sequences is explored in[9,10].

[1]INFN Sezione di Roma, Rome, Italy. [2]PhD Program in Behavioural Neuroscience, "Sapienza" University of Rome, Rome, Italy. [3]Dipartimento di Fisica, Università di Cagliari, Cagliari, Italy. [4]INFN Sezione di Cagliari, Cagliari, Italy. Correspondence and requests for materials should be addressed to C.C. (email: cristiano.capone@roma1.infn.it)

Here, we focus on sleep mediated memory association and its implications on cognitive tasks performances. We present a minimal thalamo-cortical model which, after being trained on handwritten characters in unsupervised mode, is induced to express sleep-like dynamics. We measure its effects on the classification accuracy, the structure of the synaptic matrix and firing rate distributions with findings that are consistent with experimental observations[4,8,11,12].

Slow oscillations (SO) are considered the default emergent activity of the cortical network[13] and are observed during the deepest of physiological non-REM sleep stages as an alternation between *Down states* (characterized by nearly silent neurons) and *Up states* (in which a subset of neurons goes in a high firing rate regime) occurring at a frequency in the range $[0.5, 4]$ $Hz$ (*delta* band). We set the model SO at a comparable frequency. SO activity is expected to play two complementary roles, which are separately mediated by Up states and Down states. Down states would play a purely biological function, with a lower immediate impact on cognitive performance. The role of Down states, with a majority of neurons put in a silent state for a long fraction of deep sleep time, would serve a restoration purpose, enabling periodic biological maintenance and recovery, as it happens in the whole body when at rest[14]. Our modeling and investigation is focused only on the effects mediated by the Up states dynamics. During sleep external perceptions are, at least, strongly attenuated, and the majority of the motor system is blocked[15]. For this reason in our model the local interaction between cortex and thalamus is crucial during sleep rather than contextual signal coming from other cortical modules and sensory input coming from thalamic pathways.

In order to make a biologically realistic learning protocol and to implement the role of the context in the learning phase, we took inspiration from the "organizing principle" in[16] for the Cerebral Cortex, which describes specific computational strategies implemented by the neuronal structure of Layer 5. The architecture is grounded on the separation of intra-areal and inter-areal contextual information (reaching the apical dendrites of pyramidal neurons) from the feed-forward flow of area specific information (targeting its basal synapses). Cellular mechanisms, like $Ca^{++}$ spikes, promote the detection of coincidence between contextual and specific activity. High-frequency bursts of spikes are emitted when the coincidence is detected. Relying on these observation we introduced in our model external stimuli mimiking contextual information which changes the effective firing threshold of specific subsets of neurons during the presentation of examples in the training phase. For each example, a vector of features is projected toward a cortical network by a thalamic neural network. Due to the change in the perceptual effective firing threshold, spike-timing-dependent-plasticity (STDP) creates stronger bottom-up (thalamo-cortical) and top-down (cortico-thalamic) connections between a subset of cortical and the thalamic neurons.

In our model we observe that sleep induces both the association of patterns encoding learned images belonging to the same category and a differential synaptic down-scaling. This is also reflected in a differential modulation of firing rates, producing observations similar to[4]. We observe that such effect, probably related to energetic optimization in biological networks, also has beneficial effects on the performances of our network in the image recognition task.

## Results

We tested the role and the mechanisms of the occurrence of SO in a thalamo-cortical network model which was previously trained to learn and recall images (MNIST dataset). The network model included thalamic relay (*tc*) and reticular (*re*) neurons in the thalamus, as well as pyramidal neurons (*cx*) and inhibitory interneurons (*in*) in the cortex (Fig. 1A, see Methods) following a standard minimal structure for thalamico-cortical models[17].

Figure 1B shows an example of activity time-course in the *cx* and *tc* populations during the training phase, the retrieval phase, and the early stage of sleep phase.

### Training and pre-sleep retrieval.
During the training, 9 different images were presented to the network, in a first set of runs: 3 instances for each class of digit, for a total of 3 different classes (e.g. 0, 1, 2). In a second group of runs, 30 examples per digit constituted the training set. For each image an external stimulus (*contextual signal*) induced a different subset of *cx* neurons to code for that specific image, with STDP shaping the intra-cortical, the thalamo-cortical and the cortico-thalamic connectivity. In order to adopt the prescription in[16] the parameters are set to make the cortical neurons fire during the training only if they receive both sensory and contextual stimuli. This training procedure works even in the extreme case where only one neuron is used to code each digit example. However for one or a few neurons, self sustained oscillations would not be well defined. For this reason we chose a population of 20 cortical neurons for each newly presented example. The precise number of such neurons is not a critical factor as we will discuss later.

After the training, images were presented again (retrieval phase) without the external stimulus. The population of neuron responding to the image were the same as the one selected in the training phase by the external stimulus, demonstrating the success of the retrieval.

### Induction of slow-oscillation.
A non-specific stimulus at low firing rate was provided to cortical neurons only, while model parameters are modulated (see Methods for details) eliciting the spontaneous occurrence of cortically generated Up states and of thalamo-cortical SO (see Fig. 1B).

We relied on the framework of Mean Field theory to obtain a model displaying different dynamical regimes. An oscillatory regime, closely resembling Slow Oscillations observed in deep sleep and anesthetized states can be induced by introducing a relatively strong recurrent excitation and spike frequency adaptation[18,19]. We tuned the parameters of the network to make it display SO frequency (between $0.25\,Hz$ and $1.5\,Hz$) and Up state durations (a few hundred of milliseconds) comparable with experimental observations in deep sleep recordings[20,21].

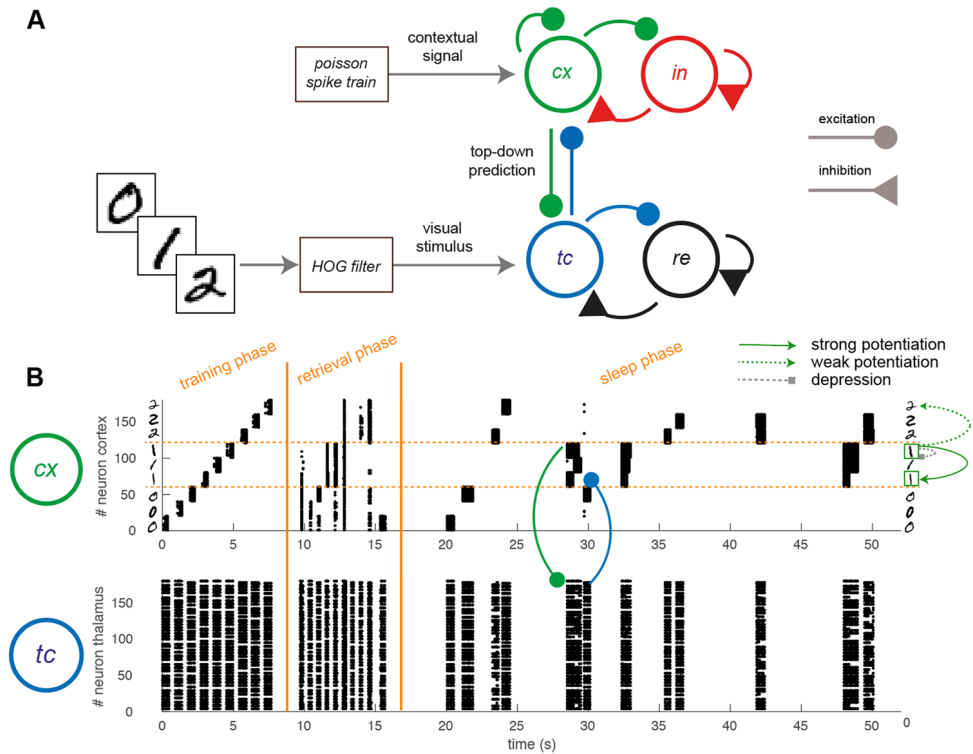

**Figure 1.** Thalamo-cortical model and protocol description. (**A**) Sketch of the structure of the simplified thalamo-cortical model considered, which is composed of an excitatory and an inhibitory population both for the cortex (*cx*, *in*) and for the thalamus (*tc*, *re*). Connectivity structure is represented by solid lines. The visual input is fed into the model through the thalamic population, mimicking the biological visual pathways. In the training phase a lateral stimulus enhances a specific subset of *cx* neurons to preferentially represent the stimulus. (**B**) Activity produced during training phase, pre-sleep retrieval and the first $40s$ of SO activity in the *cx* (top) and *tc* (bottom) populations. Only first 180 neurons in *tc* population are shown for visual purposes. In the training 3 instances of 3 classes of digits (0,1,2) are learned by the network. In the replay during sleep, thalamo-cortical connections promotes the activation of neurons coding for similar patterns of activity, causing the potentiation of cortico-cortical connections between neurons representing digits of the same class. A general depression reduces the largest synaptic weights. Post SO retrieval is not shown.

The activity (Fig. 1B) during the initial stages of simulated SO displays that Up states are independently sustained by populations coding for different memorized images. However, thanks to the cortico-thalamo-cortical pathways each population tends to recruit other populations sharing similar thalamic representations. Indeed when a population initiates an Up states, it activates thalamic patterns similar to the one responsible for the activation of the population itself during the retrieval phase. This is what we call *top-down prediction*. The thalamus, in turn, activates all the other populations which coded for a similar thalamic input. Across the sleep period, thanks to the cortico-cortical plasticity, the co-activation of populations coding for similar *predicted* synaptic input becomes a more and more prominent feature.

### Effects on the synaptic matrix of slow-oscillations.

We first considered the case with the training set composed of 3 classes and 3 examples per class. After the training stage the system undergoes a $600s$ period of sleep. During this stage the activation of groups of neurons associated to different training examples induces, through STDP, not only a synaptic pruning but also the creation of stronger synapses between groups of neurons that share enough commonality in the features they received during the training.

This effect can be noticed comparing the structure of the synaptic matrix before and after sleep as in Fig. 2A,B, that reports the change from the initial flat structure (which reflects the individual training examples) towards a hierarchical structure (embedding the categories of learned digits).

Indeed we can observe that novel synapses are created among examples in the same digit category (Fig. 2A-left). At the same time the system undergoes a down-scaling of the strongest synapses, those linking neurons coding for the same image (Fig. 2A-right). The differential effect on synapses can also be observed in Fig. 2C, where the histogram of synaptic weights after-sleep is reported. The synapses between patterns encoding for different learned images of the "same class" and those between "different classes" (respectively orange and green distribution in Fig. 2C) are originally drawn from the same distribution by definition (see Methods), while after sleep they are clearly differentiated.

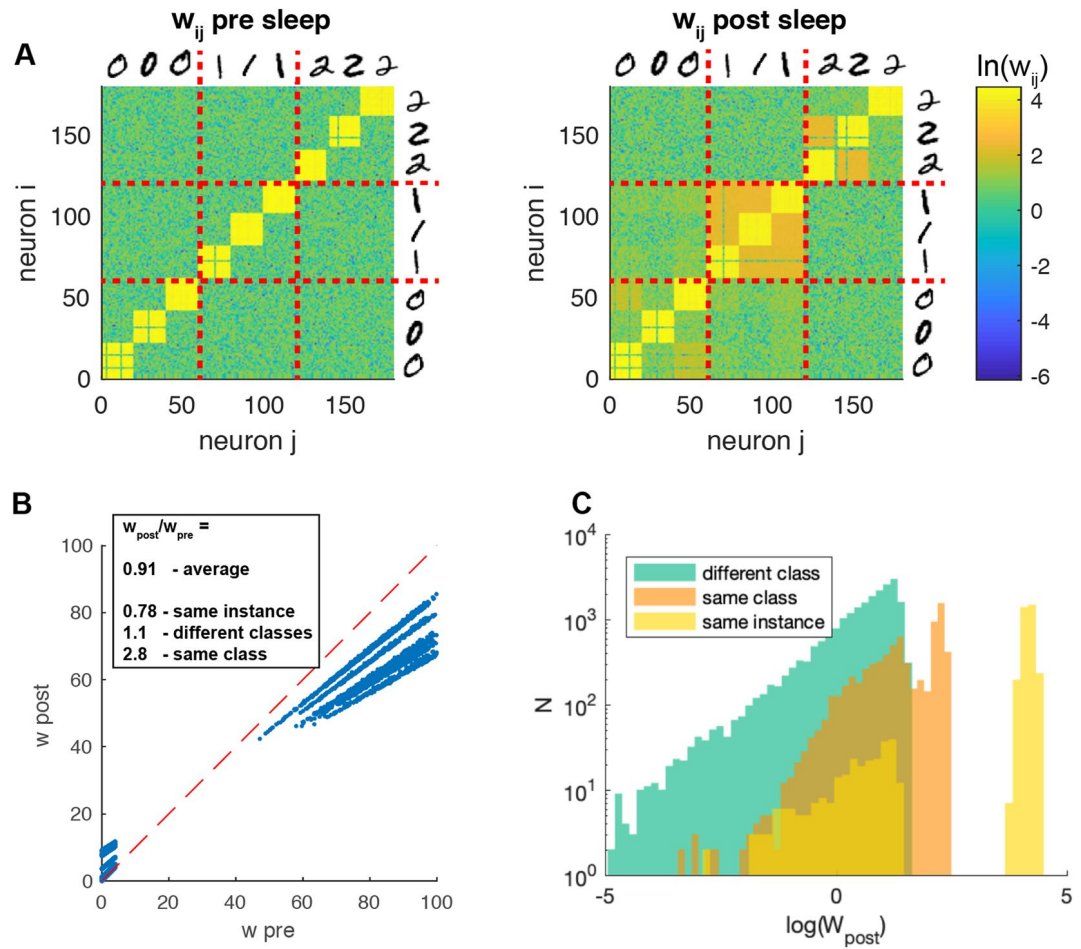

**Figure 2.** SO effects on connectivity structure. (**A**) Synaptic weights matrix of the recurrent connectivity of *cx* population, before (left) and after (right) the occurrence of sleep-like activity. The yellow squares represent high weights emerged between neurons encoding the visual input related to the same object (single instance of 0, 1, 2 … image). Red solid lines separate the neurons encoding visual inputs related to different classes of objects (0, 1, 2 …). (**B**) Scatter-plot of the same synaptic weights before and after sleep. (**C**) Synaptic weights after sleep, separated in three groups, synapses between neurons encoding the same object (yellow), the same class (but not the same object, orange) and different classes (green).

**Effects on the post-sleep activity.** The change in structure of the synaptic weights matrix modifies the activity expressed by the network in the retrieval phase. This can be appreciated looking at the difference of the correlations between groups of neurons before and after sleep (see Fig. 3A).

Figure 3B reports the difference among correlations evaluated after and before sleep. It shows decorrelation (blue squares) of populations encoding different classes, and correlation (red regions) of the ones coding the same class. Such information is reported also in Fig. 3C, showing the correlation changes for populations in the same class (blue) and in different classes (green). Such effect might provide benefits in retrieval and classification tasks.

The consistency of such result is testified by Fig. 4. There, the same simulation is performed for different training sets (different examples of 0,1,2 digits). All simulations show that the synapses between neurons in the same class are the more reinforced (Fig. 4A, orange versus green) and that their internal representation has an increased correlation (Fig. 4B, blue versus green).

**Mechanistic interpretation.** We propose that such effect is due to the interplay between cortico-thalamic predictions and thalamo-cortical connections. In other words when a group of neurons undergoes an Up state it formulates a prediction in the thalamus by activating a thalamic pattern similar to the one received during training. In turn, the thalamus projects to the cortex and activates those populations trained for similar input patterns. This mechanism promotes the connections between populations of neurons coding for the images of the same class through STDP. To prove this, we reproduced the same experiment switching off the cortico-thalamic prediction and reported the result in Fig. 4C,D. It is evident that there is no sign of the preferential association observed in the control condition, nor in the synaptic structure (Fig. 4C), neither in the internal representation (Fig. 4D).

To demonstrate the specific role of SO activity, we repeated the same experiment replacing the sleep like activity with an awake like asynchronous activity. We set the same adaptation strength $b$ and $w_{in \rightarrow cx}$ used during training and retrieval. In this case we did not observe significant changes in the synaptic matrix structure, with absence of synaptic down-scaling (homeostasis) and no association of memories. This is due both to the absence of the specific spatio-temporal structure of the activity that characterizes the SO state and the lower firing rate (indeed asynchronous state has a lower firing rate than Up states).

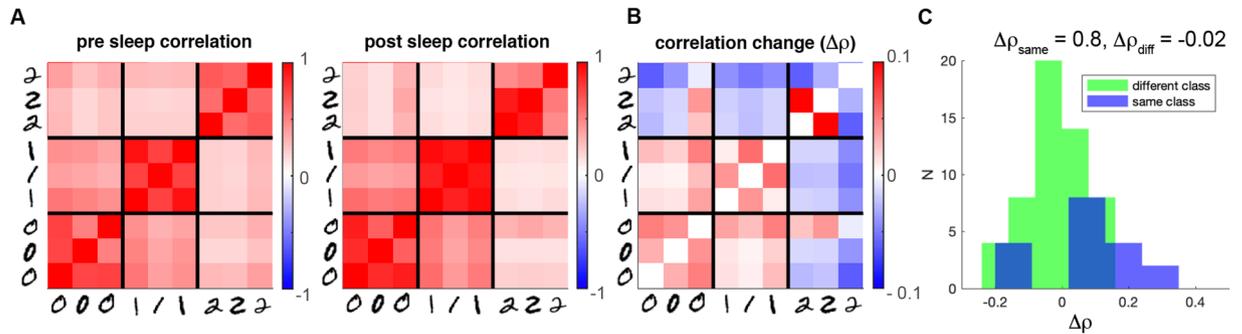

**Figure 3.** SO effects on internal representation. (**A**) Activity correlation between all pairs of populations representing the single images before (left) and after (right) sleep. (**B**) Correlation difference between after and before sleep. (**C**) Histogram of correlation differences for populations encoding the same class (blue) and different classes (green).

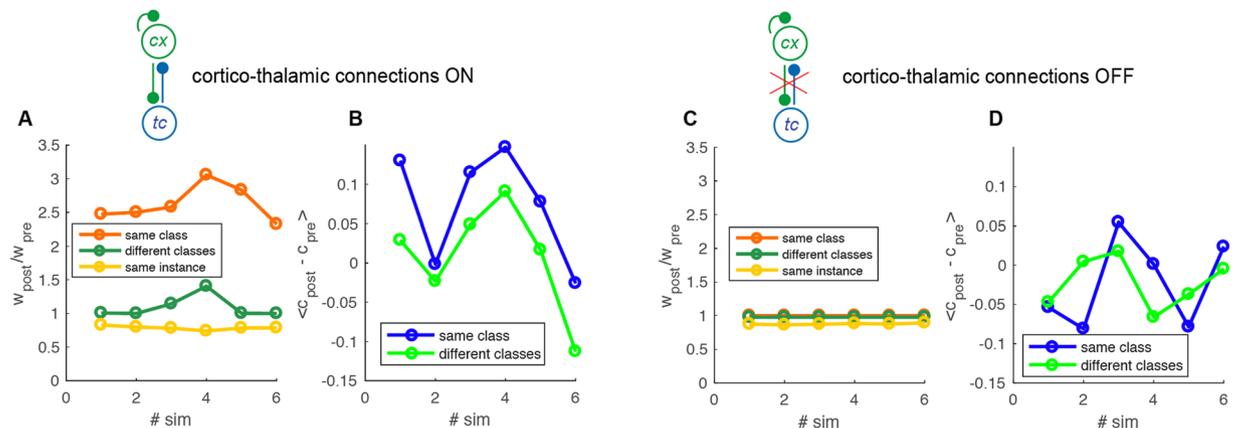

**Figure 4.** Analysis of populations: synaptic weights and comparison between correlations with and without *cortico-thalamic predictions*. (**A**) Average ratio between weights post- and pre- sleep for each simulation (top). The different categories are separated in different colors. Yellow: synapses connecting neurons coding for the same image, orange: different image of the same class of digits and green: different classes. (**B**) The average change in correlation between post- and pre- sleep for each simulation (top) and histogram of the distribution over all the simulations (n = 6, bottom). Blue: same class, green: different classes. (**C,D**) as in A-B but in absence of cortico-thalamic connections.

**Post-sleep improvement in a classification task.** Finally, we evaluated the effect of sleep-like activity on the performance of a set of classification task trials. Networks were exposed to example and test images drawn from all the ten classes of digits in the MNIST dataset. Each simulation trial used a different test set of 250 images, and a set of 3 training examples per digit (30 training instances in total, also randomly extracted for each classification trial). Each network was exposed to the training examples using the same protocol discussed above for the simpler retrieval task (see Methods for details). For each test image, the classification was determined looking for the class of the neuron responding with the higher firing rate. We note that class labels were used only during classification and not during the training that was completely unsupervised.

We observed a net increase in the classification accuracy across the sleep period. Figure 5A reports in blue the time course of accuracy increase as a function of the sleep time. After $3000s$ of sleep, the improvement was on average $6.0\% \pm 0.5\%$ (accuracy rose from $58.0\%$ to $64.0\%$, average performed over 24 simulations). In absence of thalamic feedback the improvement is significantly lower (Fig. 5A red line), proving that the memory association due to the cortico-thalamo-cortical interaction is beneficial to performance in a classification task.

Figure 5B reports the average weights evolution as a function of sleep time. Synapses between groups of neurons encoding for different instances of the same digit class (yellow solid line) were on average strongly potentiated, much more than the ones connecting training examples belonging to different classes (green solid line). Synapses interconnecting neurons representing individual training instances were down-scaled (orange solid line). When the thalamo-cortical feedback was switched off (same colors, dashed lines) this differential effect did not happen. This is the same effects already reported for the case with 9 training examples (Fig. 2, simpler retrieval task). We verified that the same qualitative results are obtained also for the cases in which each example is coded by 10, 15 and 25 cortical neurons (not shown).

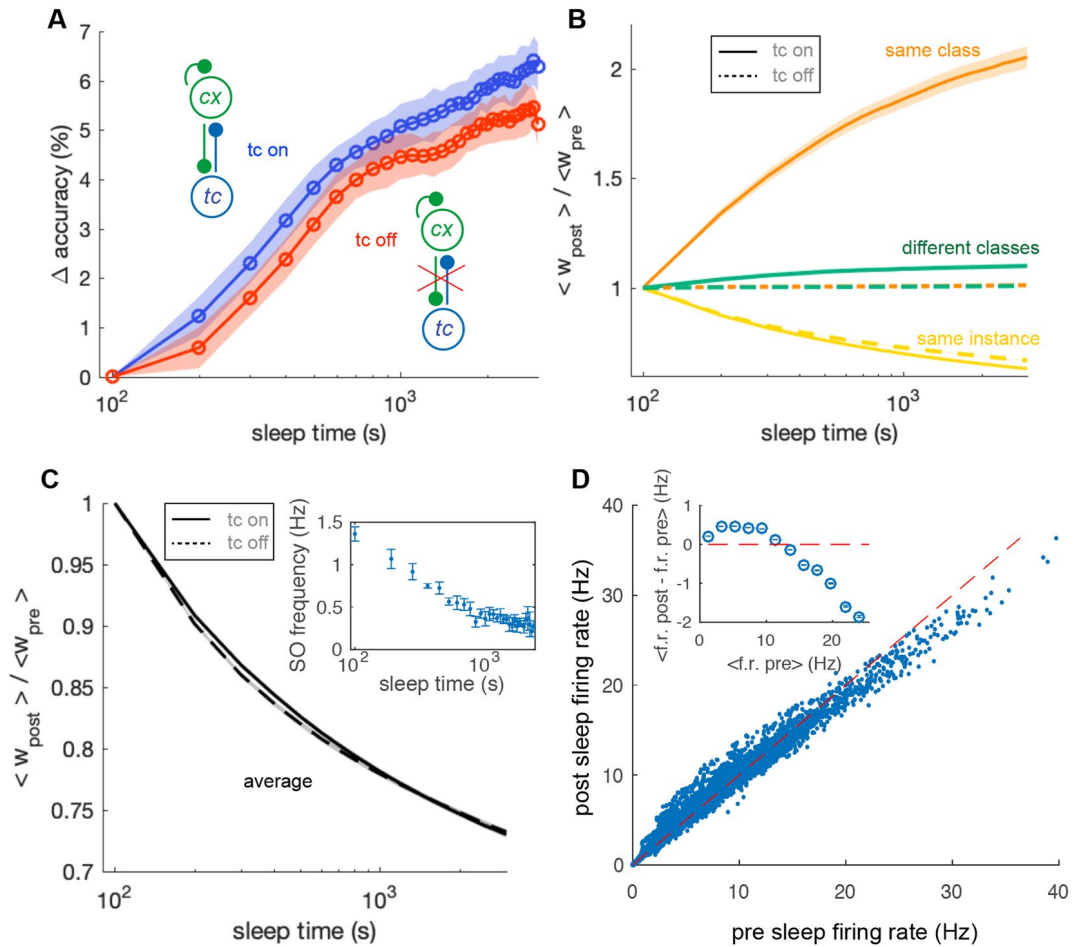

**Figure 5.** Sleep effects on a classification task. (**A**) Change in classification accuracy across over 30 sleep epochs (100s each). Blue and red are respectively the conditions in which thalamus is on and off. The improvement in accuracy is averaged over 30 simulation trials. SEM is reported in the shading. (**B**) Average synaptic potentiation and depression over 30 sleep epochs. The colors indicate connections between neurons coding the same instance (yellow), different instances of the same class (green) and instances of different classes (orange). Dashed and solid lines represent the comparison between the conditions in which thalamus is on and off. (**C**) Average synaptic depression over all the synapses. (inset) Average decrease of SO frequency across sleep time, average over 4 simulations. (**D**) Scatter of single neurons activity in 8 simulations averaged over time in a classification task before and after 3000s of sleep. Inset, average difference of activity after and before sleep as a function of activity before sleep.

Such differential mechanism occurred together with a general synaptic depression (see Fig. 5C, thalamus on and thalamus off in solid and dashed line respectively). We notice that the average down-scaling is very similar for simulations executed in absence and presence of thalamic feed-back, allowing for a fair comparison of the two conditions. As a consequence of such general synaptic depression the SO frequency decreases over time (see Fig. 5C, inset) consistently with experimental observations[12].

We observe that an optimal range of SO frequencies is important to obtain the reported results. Indeed an extremely low Up state occurrence would make weaker the specific cortico-thalamo-cortical association. On the other hand a very high Up state occurrence frequency would increase the probability to randomly associate different classes of digits. Both scenarios would impair the positive effect of the sleep period on network performances.

Firing rates also underwent a differential modulation. Neurons with pre-sleep low activity became more active after sleep (and *vice versa*). Figure 5D displays a scatter of time averaged single neuron activity during the classification task, before and after the 3000s sleep period (data are drawn from 8 simulations). The distribution of individual firing rates is rotated respect to the bisector line (red dashed line). The average difference of activity after and before sleep is positive for low values of pre-sleep firing rates and negative for high pre-sleep rates (see inset of Fig. 5D). This prediction is similar to what observed in[4], strengthening the biological plausibility of our model.

## Discussion

We propose a minimal thalamo-cortical model that classifies images drawn from the MNIST set of handwritten digits. During a training phase, spike-timing-dependent-plasticity (STDP) sculptures a pre-sleep synaptic matrix and creates top-down synapses toward the thalamus. Then, the network is induced to produce deep-sleep-like slow oscillations (SO)[19,22], while being disconnected from sensory and lateral stimuli and driven by its internal activity. During sleep Up states, thalamic cells are activated by top-down predictive stimuli produced by cortical groups of neurons and respond with a forward feedback, recruiting other cortical neurons.

In our model it was of utmost importance to obtain a biologically accurate reproduction of the association/coincidence mechanism described in[16] between sensory (from thalamus) and contextual stimuli (from external poisson processes). To do this properly would require a dedicated spiking neuronal model. The implementation of such model in neural simulation engines is a current topic of development for the community. In this study we approximated the coincidence mechanism through a careful subthreshold setting of both the contextual and sensory inputs impinging in cortical neurons. In the training phase the neuron only fired whether both contextual and sensory input was received. Our study proves the goodness of the coincidence mechanism for a fast learning on few examples only presented once, providing an additional motivation for the development of a dedicated model.

One delicate point is that the contextual signal facilitates the learning of each example in a different group of neurons. Even if mixed selectivity neurons are fundamental in neural coding[23], and are indeed created in our simulation during sleep, this work adopted this disjoint representation of the contextual facilitation signal, also with the purpose of simplifying the understanding of the association mechanism itself. However we note that, in first approximation, this mechanism would not be unreasonable in the cortex in several circumstances, for at least the following reasons. First, we are often exposed to individual examples of objects embedded in different contexts. A second argument in favour is that the internal state of the brain that dominates the creation of the contextual signal is never the same, even if external conditions are similar. Moreover, the coding for each internal state is sparse and long-range connections bringing contextual information are themselves sparse. Contextual signals with low degree of overlap are therefore plausible. The sparse selectivity of contextual signal facilitates the creation of a wide, orthogonal internal representation of individual examples, but lacks of generalization of low levels feature. Bottom-up sensory stimuli of examples belonging to the same class should present a high degree of overlap. We demonstrated that sleep could help overcoming this problem, inducing specific association of examples sharing commonality in low level features. This effect might be beneficial also to machine learning (ML) algorithms which currently lack this generalization feature and is relevant for cerebral neural networks.

We observed differences in the synaptic structure and in the activity expressed before and after the sleep-like period. Indeed SO induced two effects that are both beneficial and biologically plausible. The first is a reduction in the strength of synapses inside neural populations created by the training on specific examples, the second is a simultaneous increase of synapses that associate examples of characters belonging to the same category of digits. The final structure of the synaptic matrix is more complex and hierarchically organized. Moreover, the correlation of groups of neurons during post-sleep retrieval activity reflects the hierarchical structure supported by the underlying synaptic matrix, with stronger cooperation between groups of neurons trained by different examples belonging to the same class. Also, during the sleep phase, the network displays a rich internal dynamics that evolves across sleep time. Indeed, the composition of neuronal groups recruited during the Up states changes sensibly between the initial and the final stages of sleep.

We also investigated the benefits of such effects on the image classification task on the MNIST dataset, finding an increase in classification performance. This improvement is due to both the described mechanisms. First the synaptic down-scaling prevent the system from signal saturation, improving the sensibility of the network. Second, the association due to thalamo-cortical interaction brings different groups of neurons to cooperate, leading to a more reliable result during the classification task.

Interestingly, our simulations produce dynamical change predictions that are coherent with recent biological observations. First we observed that the SO frequency decreases during the sleep period. This is due to the asymmetric STDP we used, and is consistent with experimental observations, where a decrease in SO frequency is observed over night[12]. This also serves as a protection mechanism, to stop Up state mediated associations before a possible catastrophic divergence to a fully connected network. Second, we found that neurons with low levels of activity before sleep increased their firing rate after sleep and vice-versa, which is very similar to the global effect of sleep observed in[4], strengthening the biological relevance of our model.

We stress that SO activity is fundamental to achieve the results discussed in this paper, while asynchronous state lacks the spatio-temporal structure suited to provide the memory association and the synaptic down-scaling that we observed. Up states did not simultaneously occur in all the groups of neurons, but only populations coding for images in the same class were likely to co-activate. The high level of simultaneous firing rates induced stronger synapses between group of neurons coding for examples belonging to the same class, while populations of different classes remained largely unaffected. Also, the high firing rate promoted a generalized synaptic down-scaling thanks to the asymmetric STDP rule. Such specific synaptic organizations would not be possible during asynchronous activity.

One of the main peculiarities of this work is that, despite being biologically oriented, it investigates the effects of SO activity on networks that perform ML relevant tasks. This follows a path of works in which different tasks (such as image recognition and categorization) are performed in networks which are constrained to use biologically plausible mechanisms such as STDP plasticity and spike-based signal transmission[24–27]. In this framework the impact of the interaction between sleep-like activity and plasticity on the network performances and dynamics has never been addressed. Indeed, in every day life humans and machines might be exposed, in different spatial, temporal or social context, to instances of external objects belonging to the same class. Still, they could not immediately realize that they are correlated, e.g. because driven by the urgent need to solve the current task.

The association of examples sharing similar low-level features could be performed after the learning, e.g. during sleep. With this work, we demonstrated a first mechanism by which sleep can be beneficial for both human and artificial intelligence.

We acknowledge that in our model there are some strong assumptions, that make the network stereotyped and still far from biological reality. For example cortical neuron are initially set to encode trained images in a way that is completely disjoint using orthogonality in the contextual signal. Despite this is not completely unreasonable in some extent (as discussed above), it is known that mixed selectivity neurons are fundamental in neural coding[23] everywhere, and indeed are created in our model during sleep induced association. We plan to include mixed selectivity in the contextual signal in future work.

Also, we plan to perform large scale simulations of Slow Wave Activity (SWA) on cortical areas with biologically plausible long-range lateral connections, and columnar organization (see Pastorelli *et al*.[28]), this way supporting the integration of contextual lateral information in cortical model with retino-topical organization. We speculate that the introduction of such spatial extension might allow our model to express the sub-state specificity observed in[4] and support the different features of Slow Oscillations and Micro Arousal states. On the long term, such a detailed approach should enable a comparison of simulation results with experimental measures of slow waves propagation patterns and transitions across different stages of sleep.

In the framework we proposed it will be interesting to test and compare different hypotheses of sleep functions, some of which are not yet unified in a coherent framework[1,4]. We hope to contribute to refine these hypotheses: they could find a conciliation path according to our preliminary simulation results, that point to the possibility of having SO promoting both a hierarchical association of memories and differential synaptic homeostasis mechanisms.

Our approach is also promising in the direction of biologically plausible ML, and in the next future we aim to obtain networks with superior recognition accuracy, and test out network on more complicated image datasets.

In[15] the authors discuss the role of sleep in relation with the Integrated Information Theory (IIT) of consciousness. We remark that the groups of neurons created by the pre-sleep training resemble the elementary mechanisms of the conceptual framework described in IIT[29]. The increase in complexity among the groups resulting from sleep, clearly visible in the final synaptic matrix structure, could be viewed as a step towards the creation of higher order mechanisms. This should be associated to changes in the distribution of probabilities of the states accessible to the system itself and on a change of its causal power. This model might constitute a simplified biological computational setup to investigate this kind of conceptual framework.

## Methods

**Network architecture.** The network, which is designed to be a minimal thalamo-cortical model[17] (see Fig. 1), is composed of two populations of cortical neurons (one excitatory *cx* and one inhibitory *in*) and two thalamic populations (one excitatory *tc* and one inhibitory *re*).

The role of cortical inhibitory neurons in our model is to provide a shared inhibition supporting a winner-take-all mechanism. Indeed when a part of the whole network responds to a specific input, the rest of the network is inhibited. A classical choice to approximate a biological cortical network is to set a 4:1 ratio between excitatory and inhibitory neurons. Keeping a fixed excitatory:inhibitory ratio is not critical for the model here presented. In our runs, the ratio varied from 4:1 to 1:1 because for simplicity we kept the number of inhibitory fixed in all simulations, while we increased the number of excitatory with to the number of training examples.

The connection probability is $p = 1.0$ for the populations connected by the arrows in Fig. 1. The untrained synaptic weights are $w_{in \to cx} = -4, w_{cx \to in} = 60, w_{tc \to re} = 10, w_{re \to tc} = -10, w_{in \to in} = -1, w_{re \to re} = -1$. A subset of synapses ($w_{cx \to cx}, w_{cx \to tc}, w_{tc \to cx}$) is plastic: their initial value and plasticity rules are specified later-on.

The *contextual signal* is a Poissonian train of spikes which mimics a contextual signal coming from other brain areas and selectively facilitates neurons to learn new stimuli. The *top-down prediction* is the signal flowing through $cx \to tc$ connections, predicting the thalamic configuration which activated a specific cortical activity pattern.

**Pre-processing of visual input.** Images are pre-processed through the application of the histogram of oriented gradients (HOG) algorithm. The size of original images is $28 \times 28$ pixel. Histograms are computed using cells of $14 \times 14$ size that are applied on the images using a striding step of 7 pixels, resulting in 9 histograms per image. Each cell provides an histogram with 9 bins, each bin assuming a real value. Each bin is then transformed into a vector of 4 mutually exclusive truth values. In summary, each image has been transformed in a vector of 324 binary features. The resulting visual input is presented to the network through the thalamus, where each feature provides a binary input to a different thalamic cell. Each cell receives a Poisson spike train with average firing rate that is $30\,kHz$ only when the element of the feature vector is 1. The specific number of thalamic neurons used in the model is related to the specific pre-processing algorithm and the number of levels used to code the pre-processing output.

**Network size and training set.** The number of thalamic (*tc*) neurons is the same as the dimension of the feature vector produced by the pre-processing of visual input ($N_{th} = 324$). The cortical population is composed of groups of 20 neurons for each image in the training set. In a first set of runs, the training set is composed of 9 images, with 3 different examples for 3 different digits (0, 1, 2). In this case the number of cortical neurons is $N_{cx} = 180$. In a second run, the number of training examples has been raised to 30 images (3 examples per digit, $N_{cx} = 600$). The number of inhibitory neurons are $N_{in} = 200$ and $N_{re} = 200$ (thalamic inhibition).

**Training algorithm and retrieval.** Every time a new training image is presented to the network through the thalamic pathway, the facilitation signal coming from the contextual signal provides a $2\,kHz$ Poisson spike train to a different set of 20 neurons, inducing the group to encode for that specific input stimulus (see the Discussion section for details about this choice). Additionally a 10 *kHz* Poisson spike train is provided to inhibitory neurons (*in*) to prevent already trained neurons to respond to new stimuli in the training phase. Synaptic weights for Poisson inputs are $w_{tc}^{poiss} = 8$, $w_{cx}^{poiss} = 15$ and $w_{in}^{poiss} = 5$. The learning mechanism is allowed by symmetric ($\alpha = 1.0$) spike-timing-dependent-plasticity (STDP) present in the $cx \rightarrow cx$, $cx \rightarrow tc$ and $tc \rightarrow cx$ connections which shapes the weights structure. The maximum synaptic weight (see STDP eq. 4) are respectively $w_{cx \rightarrow cx}^{max} = 150$, $w_{cx \rightarrow tc}^{max} = 130$ and $w_{tc \rightarrow cx}^{max} = 5.5$. The initial values are $w_{cx \rightarrow cx}^{0} = 1$, $w_{cx \rightarrow tc}^{0} = 1$ and $w_{tc \rightarrow cx}^{0} = 1$. During the retrieval phase only the 30 *kHz* input to thalamic cell is provided, while the contextual signal is off.

**Image classification task on MNIST dataset.** Notwithstanding the aim to achieve biological plausibility, our work is also oriented to perform tasks that are relevant in the machine learning (ML) scenario. Indeed in this work we extended a model developed in a previous work, which was designed to perform an image classification task relying on a small number of examples. In that model, the *contextual signal* projected toward cortical neurons selected groups of target neurons during an incremental training of MNIST handwritten characters. The training presented 25 novel characters per training step. After each training step, we measured the growth in the recognition rate. The network reached an average recognition rate of 92% (average over 6 simulations) after the presentation of 300 example characters (i.e. 30 examples per digit).

**Slow oscillations.** After the training stage, the sleep-like thalamo-cortical spontaneous slow oscillations activity is induced for a total duration of 600*s* by providing a non-specific Poisson noise inside the cortex (700 *Hz*) and increasing the strength of SFA parameter ($b = 60$, in eq. (1)). No external stimulus is provided to *tc* cells. Also, the synaptic weights between inhibitory and excitatory neurons in the cortex is reduced to $w_{in \rightarrow cx} = -0.5$. In this stage asymmetric STDP plasticity ($\alpha = 3.0$) is active in the recurrent *cx* connectivity, inducing sleep-induced modification in the synaptic weights structure. The parameters' change to obtain the slow oscillating regime were chosen relying on mean field theory framework[18,19].

**Simulation protocol.** The simulated experimental protocol is composed of 4 phases (see Fig. 1B): the training phase, where visual patterns are learned by the network, the pre-sleep retrieval phase, in which patterns are recalled, the sleep phase and the post-sleep phase. During post-sleep, the network is either exercised on the retrieval of previous learned patterns or applied to the classification of novel examples. The network structure and dynamics are compared in the pre- and post- sleep phases.

**AdEx neurons.** The neurons in the network are adaptive exponential (adEx) point like single compartment neurons.

$$\begin{cases} \dfrac{dV(t)}{dt} = -\dfrac{V(t) - E_l}{\tau_m} + \dfrac{\Delta V}{\tau_m} e^{\left(\frac{V(t) - \theta}{\Delta V}\right)} + \dfrac{I(t, V(t))}{C} - \dfrac{W(t)}{C} \\ \dfrac{dW(t)}{dt} = -\dfrac{W(t)}{\tau_W} + b \sum_k \delta(t - t_k) + a(V(t) - E_l) \end{cases} \quad (1)$$

where the synaptic input *i* is defined as

$$I(t, V(t)) = \sum_{\alpha} g_{\alpha}^{syn}(t)(V(t) - E_{\alpha}^{syn}) \quad (2)$$

where $V(t)$ is the membrane potential of the neuron and $\alpha = e, i$ defines the excitatory (*e*) and the inhibitory (*i*) input. The population dependent parameters are: $\tau_m$ the membrane time constant, $C$ the membrane capacitance, $E_l$ the reversal potential, $\theta$ the threshold, $\Delta V$ the exponential slope parameter, $W$ the adaptation variable, $a$ and $b$ the adaptation parameters, and $g_{\alpha}^{syn}$ the synaptic conductance, defined as

$$g_{\alpha}^{syn}(t) = \sum_k \Theta(t - t_k) Q_{\alpha} \exp(-(t - t_k)/\tau_{\alpha}^{syn}) \quad (3)$$

We define the spiking time of the neuron when the membrane potential reaches the threshold $V_{spike} = \theta + 5\Delta V$. $t_k^{\alpha}$ indicates the times of pre-synaptic spikes received by the neuron from synapse type $\alpha$ with characteristic time $\tau_{\alpha}^{syn}$ and its synaptic efficacy $Q_{\alpha}$. Where non specified, the parameters are the standard defined in the "aeif_cond_alpha" neuron in the NEST simulator.

**STDP plasticity.** We used spike-timing-dependent-plasticity (STDP) which potentiates $w_{ij}$ when spike occurs earlier in neuron *j* (spike time $t_j$) than in neuron *i* (spike time $t_i$) and viceversa. We considered STDP in its multiplicative form[30] which is described by the following equation

$$\begin{cases} \Delta w_{ij} = \lambda(w^{max} - w_{ij})e^{-|t_i - t_j|/\tau} & potentiation \\ \Delta w_{ij} = -\alpha \lambda w_{ij} e^{-|t_i - t_j|/\tau} & depression \end{cases} \quad (4)$$

where $w_{max}$ is the maximum weights value, $\alpha$ is the asymmetry parameter between potentiation and depression, $\lambda$ is the learning rate and $\tau$ is the STDP timescale.

**Simulation engine.** We performed spiking simulation using the NEST simulation engine, the high-performance general purpose simulator developed by the NEST Initiative, release 2.12[31].

## References

1. Tononi, G. & Cirelli, C. Sleep and the price of plasticity: From synaptic and cellular homeostasis to memory consolidation and integration. *Neuron* **81**, 12–34, https://doi.org/10.1016/j.neuron.2013.12.025 (2014).
2. Tononi, G. *et al.* Center for sleep and consciousness - research (2018).
3. Killgore, W. D. Effects of sleep deprivation on cognition. In Kerkhof, G. A. & van Dongen, H. P. (eds) "*Human sleep and cognition*", vol. 185 of *Progress in Brain Research*, 105–129, https://doi.org/10.1016/B978-0-444-53702-7.00007-5 (Elsevier, 2010).
4. Watson, B., Levenstein, D., Greene, J., Gelinas, J. & Buzsáki, G. Network homeostasis and state dynamics of neocortical sleep. *Neuron* **90**, 839–852, https://doi.org/10.1016/j.neuron.2016.03.036 (2016).
5. Walker, M. P. & Stickgold, R. Sleep, memory, and plasticity. *Annu. Rev. Psychol.* **57**, 139–166 (2006).
6. Jadhav, S. P., Kemere, C., German, P. W. & Frank, L. M. Awake hippocampal sharp-wave ripples support spatial memory. *Science* **336**, 1454–1458 (2012).
7. Smulders, F., Kenemans, J., Jonkman, L. & Kok, A. The effects of sleep loss on task performance and the electroencephalogram in young and elderly subjects. *Biological psychology* **45**, 217–239 (1997).
8. González-Rueda, A., Pedrosa, V., Feord, R. C., Clopath, C. & Paulsen, O. Activity-dependent downscaling of subthreshold synaptic inputs during slow-wave-sleep-like activity *in vivo*. *Neuron* **97**, 1244–1252 (2018).
9. Wei, Y., Krishnan, G. P. & Bazhenov, M. Synaptic mechanisms of memory consolidation during sleep slow oscillations. *Journal of Neuroscience* **36**, 4231–4247 (2016).
10. Wei, Y., Krishnan, G. P., Komarov, M. & Bazhenov, M. Differential roles of sleep spindles and sleep slow oscillations in memory consolidation. *PLoS computational biology* **14**, e1006322 (2018).
11. de Vivo, L. *et al.* Ultrastructural evidence for synaptic scaling across the wake/sleep cycle. *Science* **355**, 507–510, https://doi.org/10.1126/science.aah5982 (2017).
12. Hobson, J. A. & Pace-Schott, E. F. The cognitive neuroscience of sleep: neuronal systems, consciousness and learning. *Nature Reviews Neuroscience* **3**, 679 (2002).
13. Sanchez-Vives, M. V., Massimini, M. & Mattia, M. Shaping the default activity pattern of the cortical network. *Neuron* **94**, 993–1001, https://doi.org/10.1016/j.neuron.2017.05.015 (2017).
14. Vyazovskiy, V. V. & Harris, K. D. Sleep and the single neuron: the role of global slow oscillations in individual cell rest. *Nature Reviews Neuroscience* **14**, 443–451, https://doi.org/10.1038/nrn3494 (2013).
15. Bucci, A. & Grasso, M. Sleep and dreaming in the predictive processing framework. In Metzinger, T. K. & Wiese, W. (eds) *Philosophy and Predictive Processing*, chap. 6, https://doi.org/10.15502/9783958573079 (MIND Group, Frankfurt am Main, 2017).
16. Larkum, M. A cellular mechanism for cortical associations: an organizing principle for the cerebral cortex. *Trends in Neurosciences* **36**, 141–151, https://doi.org/10.1016/j.tins.2012.11.006 (2013).
17. Destexhe, A. Self-sustained asynchronous irregular states and up–down states in thalamic, cortical and thalamocortical networks of nonlinear integrate-and-fire neurons. *Journal of computational neuroscience* **27**, 493 (2009).
18. Gigante, G., Mattia, M. & Del Giudice, P. Diverse population-bursting modes of adapting spiking neurons. *Physical Review Letters* **98**, 148101 (2007).
19. Capone, C. *et al.* Slow waves in cortical slices: how spontaneous activity is shaped by laminar structure. *Cerebral Cortex* 1–17 (2017).
20. Contreras, D. & Steriade, M. Cellular basis of eeg slow rhythms: a study of dynamic corticothalamic relationships. *Journal of Neuroscience* **15**, 604–622 (1995).
21. Steriade, M., Dossi, R. C. & Nunez, A. Network modulation of a slow intrinsic oscillation of cat thalamocortical neurons implicated in sleep delta waves: cortically induced synchronization and brainstem cholinergic suppression. *Journal of Neuroscience* **11**, 3200–3217 (1991).
22. Sanchez-Vives, M. & Mattia, M. Slow wave activity as the default mode of the cerebral cortex. *Arch. Ital. Biol* **152**, 147–155 (2014).
23. Rigotti, M. *et al.* The importance of mixed selectivity in complex cognitive tasks. *Nature* **497**, 585 (2013).
24. Sacramento, J., Costa, R. P., Bengio, Y. & Senn, W. Dendritic error backpropagation in deep cortical microcircuits. *arXiv preprint arXiv:1801.00062* (2017).
25. Diehl, P. U. & Cook, M. Unsupervised learning of digit recognition using spike-timing-dependent plasticity. *Frontiers in computational neuroscience* **9**, 99 (2015).
26. Cayco-Gajic, N. A., Clopath, C. & Silver, R. A. Sparse synaptic connectivity is required for decorrelation and pattern separation in feedforward networks. *Nature Communications* **8**, 1116 (2017).
27. Nicola, W. & Clopath, C. Supervised learning in spiking neural networks with force training. *Nature communications* **8**, 2208 (2017).
28. Pastorelli, E. *et al.* Gaussian and exponential lateral connectivity on distributed spiking neural network simulation. In *2018 26th Euromicro International Conference on Parallel, Distributed and Network-based Processing (PDP)*, vol. 00, 658–665, https://doi.org/10.1109/PDP2018.2018.00110 (2018).
29. Tononi, G. & Koch, C. Consciousness: here, there and everywhere? *Philosophical Transactions of the Royal Society B: Biological Sciences* **370**, https://doi.org/10.1098/rstb.2014.0167 (2015)..
30. Gütig, R., Aharonov, R., Rotter, S. & Sompolinsky, H. Learning input correlations through nonlinear temporally asymmetric hebbian plasticity. *Journal of Neuroscience* **23**, 3697–3714 (2003).
31. Kunkel, S. *et al.* Nest 2.12.0, https://doi.org/10.5281/zenodo.259534 (2017).

## Acknowledgements

This work has been supported by the European Union Horizon 2020 Research and Innovation program under the FET Flagship Human Brain Project (SGA2 grant agreement SGA2 n. 785907), System and Cognitive Neuroscience subproject, WaveScalES experiment. This work is part of the activities performed by the INFN APE Parallel/Distributed Computing laboratory, and we are grateful to the members of the APE lab for their strenuous support.

## Author Contributions

C.C., B.G. and P.P. conceived the experiment, C.C. and E.P. conducted the experiment, C.C. and P.P. analyzed the results. All authors wrote and reviewed the manuscript.

## Additional Information

**Competing Interests:** The authors declare no competing interests.

**Publisher's note:** Springer Nature remains neutral with regard to jurisdictional claims in published maps and institutional affiliations.